\documentclass[12pt]{iopart}
\usepackage[english]{babel}
\usepackage{graphicx}
\usepackage{bbold}
\usepackage{color}
\usepackage{ulem}
\usepackage{gensymb}
\usepackage{latexsym}
\usepackage{amssymb}
\usepackage{multirow}
\usepackage{booktabs}
\usepackage{tablefootnote}

\usepackage{cleveref}
\usepackage{tabularx}
\usepackage{atbegshi}

\usepackage{scrextend}
\usepackage{hyperref}
\newcommand {\mo}{MoS$_2$}

\begin{document}

\title{Narrow photoluminescence peak of epitaxial MoS$_2$ on graphene/Ir(111)}

\author{Niels Ehlen$^1$, Joshua Hall$^1$, Boris V. Senkovskiy$^1$, Martin Hell$^1$,
Jun Li$^1$, Alexander Herman$^1$, Dmitry Smirnov$^2$,  Alexander Fedorov~$^{1,2,3}$,
Vladimir Yu. Voroshnin~$^{2,3}$, Giovanni~di Santo$^4$, Luca Petaccia$^4$, Thomas
Michely $^1$, Alexander Gr\"uneis$^1$}
\address{$^1$ II. Physikalisches Institut, Universit\"at zu K\"oln, Z\"ulpicher
Stra{\ss}e 77, 50937 K\"oln, Germany}
\address{$^2$ Helmholtz-Zentrum Berlin f\"ur Materialien und Energie, Elektronenspeicherring BESSY II, Albert-Einstein-Stra{\ss}e 15, 12489 Berlin, Germany}
\address{$^3$ IFW Dresden, Helmholtzstra{\ss}e 20, 01069 Dresden, Germany}
\address{$^4$ Elettra Sincrotrone Trieste, Strada Statale 14 km 163.5, 34149 Trieste,
Italy}

\begin{abstract}
We report on the observation of photoluminescence (PL) with a narrow 18~meV peak
width from molecular beam epitaxy grown $\mathrm{MoS_2}$ on graphene/Ir(111). This
observation is explained in terms of a weak graphene-MoS$_2$ interaction that
prevents PL quenching expected for a metallic substrate. The weak
interaction of MoS$_2$ with the graphene is highlighted by angle-resolved photoemission spectroscopy and
temperature dependent Raman spectroscopy. These methods reveal that there is no hybridization
between electronic states of graphene and MoS$_2$ and a different thermal expansion
of graphene and MoS$_2$. Molecular beam epitaxy grown MoS$_2$ on graphene is therefore an
important platform for optoelectronics which allows for large area growth with
controlled properties.
\end{abstract}

\maketitle

\section{Introduction}
Following in the wake of graphene research, the optical properties of monolayer
$\mathrm{MoS_2}$ and related materials have stimulated intense research efforts over
the last years~\cite{Mak2010,Manzeli2017,Wang2017}. The MoS$_2$ monolayer can take
the form of 2H or 1T (1T') crystal structures~\cite{Chhowalla2013}, with the 2H phase
being a two-dimensional semiconductor with a direct band gap that exhibits
photoluminescence~\cite{Mak2010}. Research has shown promise for applications of
$\mathrm{MoS_2}$ as field effect transistors, electroluminescent
devices~\cite{Radisavljevic2011,Lien2018} and in the area of
spintronics~\cite{Kim2017}. However, most progress in our understanding of this
material is still based on exfoliated layers, e.g. the recently observed record
narrow luminescence of 5~meV~\cite{Cadiz2017}. Small flake size and the inherent
inability of exfoliation for scale up impedes not only scientific research using
methods where a large area film with a single orientation is needed. It also
precludes the development of $\mathrm{MoS_2}$ based electronics. 
A clean and scalable approach to \mo~and other transition metal dichalcogenide
synthesis is very low pressure chemical vapour
deposition (CVD) using a catalytically active metallic substrate to support the
decomposition of a sulphur containing precursor molecule. 
For example, simultaneous supply of Mo and H$_2$S molecules yields large islands and
even single domain monolayer coverage of \mo~on Au(111)
\cite{Gronborg2015,Bana2018}. However, the substantial interaction and hybridization
of the layer with the metallic substrate modifies the properties of the layer
substantially. This is a drawback specifically when considering potential applications in
optics. Due to the low reactivity of van der Waals substrates like graphene or
hexagonal boron nitride, neither phase pure layers nor a well defined epitaxial
relation could be realized up to now with such sulphur containing precursor
molecules \cite{Miwa2015}. Through molecular beam epitaxy (MBE) using elemental
sulphur -- supplied e.g. from a valved sulphur cracker cell or from a Knudsen cell
releasing elemental sulphur out of a compound like FeS$_2$ -- phase pure and
epitaxial transition metal disulfide layers could be grown even on van der Waals
substrates to which they are only weakly bonded~\cite{Fu17,Hall2018}.

However, a complete spectroscopic characterization of such heterostructures is missing so far despite the fundamental interest in MoS$_2$ on graphene ($\mathrm{MoS_2/Gr}$) e.g. as a photodetector~\cite{Froehlicher2018}. Moreover, none of the above mentioned works on MBE grown $\mathrm{MoS_2}$ reported optical (photoluminescence or Raman) characterization of the material. This is surprising because optical methods are a main tool for the investigation of exfoliated $\mathrm{MoS_2}$~\cite{Wang2017}. The lack of optical spectroscopy characterization for MBE grown MoS$_2$ might be explained by the fact that these methods are less prevalent in the MBE community.

The present manuscript addresses these points and, besides structural investigation,
investigates MBE-grown $\mathrm{MoS_2}$ spectroscopically using X-ray photoemission
spectroscopy (XPS), angle-resolved photoemission spectroscopy (ARPES) and optical
(Raman and luminescence) methods. For the monolayer islands of
$\mathrm{MoS_2}$ epitaxially grown on a closed layer of graphene on Ir(111), as seen
by scanning tunneling microscopy (STM) and low energy electron diffraction (LEED), the band structure measured by ARPES highlights the absence of any
hybridization between $\mathrm{MoS_2}$ and graphene. Our results reveal that the
photoluminescence (PL) of $\mathrm{MoS_2}$/Gr/Ir(111) is present despite the
metallic substrate. We compare the optical bandgap obtained from PL measurements of the pristine
$\mathrm{MoS_2}$/Gr/Ir(111) system to the energy separation between valence and
conduction bands of the lithium (Li) doped system that we measured using ARPES. By careful analysis of this data and taking into account the doping induced bandgap renormalization, we estimate an exciton binding energy of 480meV. The temperature dependence of the bond lengths in graphene and
$\mathrm{MoS_2}$ is probed using Raman spectroscopy. We find that the lattice
expansion of graphene and $\mathrm{MoS_2}$ behave completely different. Graphene's
lattice expansion is dictated by the underlying Ir. The layer of $\mathrm{MoS_2}$, which is not
in direct contact to the Ir(111), roughly follows the lattice expansion expected for
freestanding MoS$_2$. Our findings introduce MBE grown $\mathrm{MoS_2}$/Gr as a
highly ordered, epitaxial heterostructure with a sharp optical emission that can be
grown in large scale.

%\cite{Amani2015,Borer1971,Cadiz2017,Froehlicher2018,Gaudreau2013,GrÃƒÂ¸nborg2015,Hill2015,Korn2011,Malic2014,Menendez1984,Miwa2015,Mohiuddin2009,Najmaei2013,Park2018a,Postmus1968,Rice2013,Ryou2016,Schuller2013,Su2014,Wang2017,Yang2016,Zhang2014}
\begin{figure}[htb!]
\centering
\includegraphics[width=\textwidth]{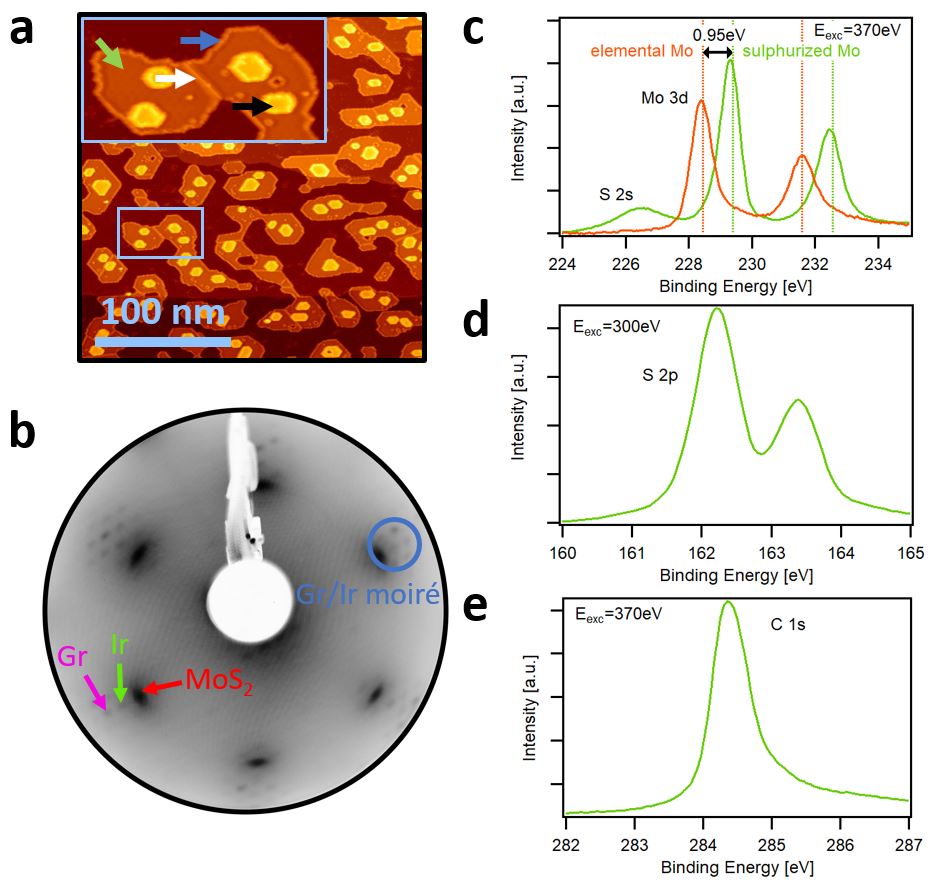}
\caption{ (a)~Large scale STM topograph of MoS$_{2}$ islands on fully covered
Gr/Ir(111) grown in two successive growth cycles (cf. Methods) and subsequently
imaged in situ. The inset shows a zoomed in part of the image marked with a blue box. The arrows in the inset mark different features of the topograph. green: monolayer MoS$_2$ island, black: bilayer MoS$_2$ island, white: mirror twin boundary, blue: metallic edge state. Image information: Tunneling voltage -1.5\,V, tunneling current 20 pA, image size 250 $\times$
250 nm$^2$. (b)~Corresponding LEED pattern at 86 eV. (c) XPS spectra of the Mo~3d peak. The doublet peak is split into
$\mathrm{Mo~3d_{5/2}}$ and $\mathrm{Mo~3d_{3/2}}$ components. The red color traces
show the peak for elemental Mo, green color for $\mathrm{MoS_2}$ which is shifted by
0.95~eV to a higher binding energy. The additional peak in the green spectrum around
226.5~eV binding energy is due to the S~2s core level. The spectra were recorded
using an excitation energy of 370~eV. (d) S~2p doublet peak from the grown
$\mathrm{MoS_2}$ using 300~eV excitation energy. (e) XPS Spectra of the C~1s peak using an excitation energy of 370~eV. }\label{fig:XPS}
\end{figure}

\begin{figure}[htb!]
\centering
\includegraphics[width=\textwidth]{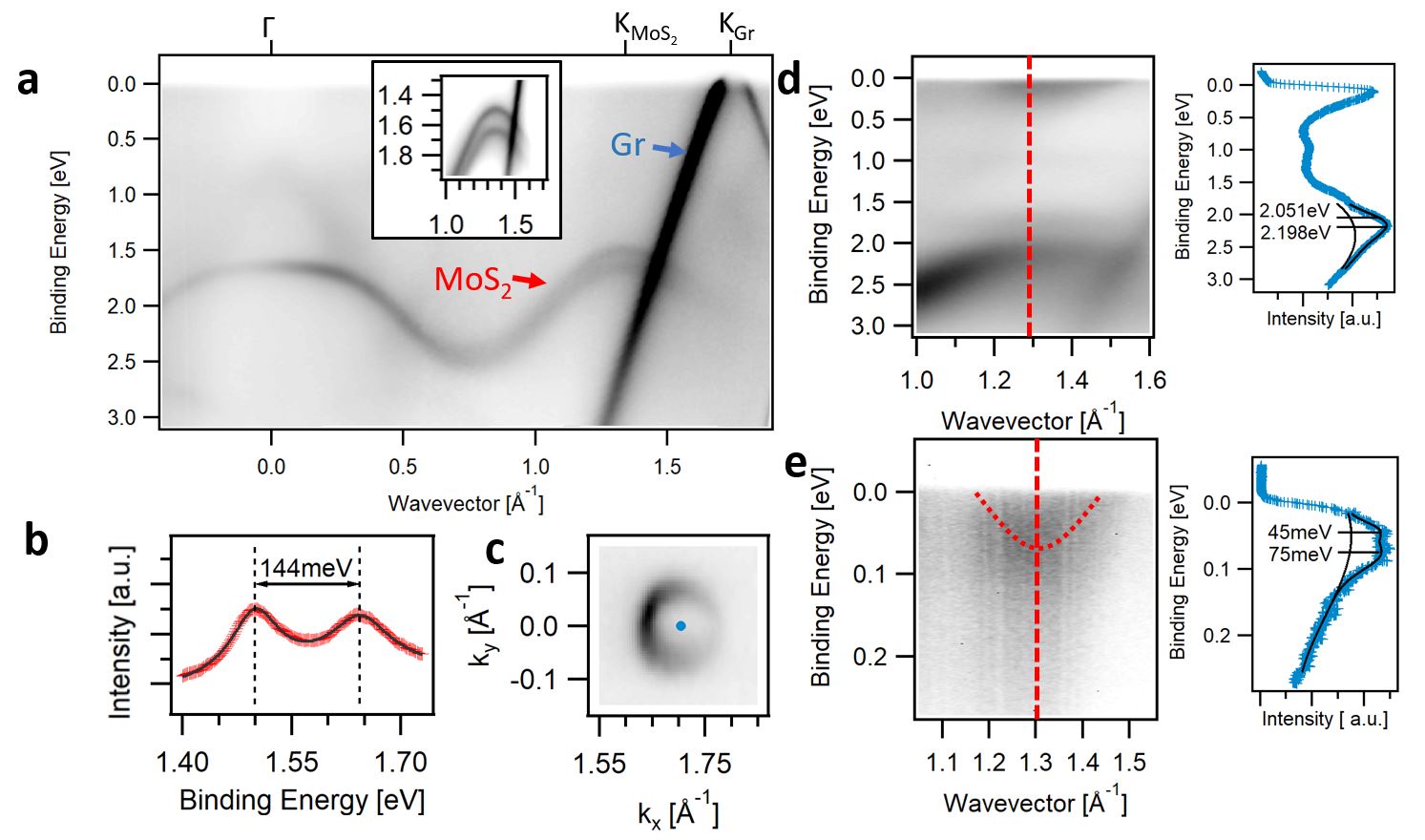}
\caption{(a) ARPES spectra of MoS$_2$/Gr/Ir(111) taken with p-polarized light at $h\nu=$31~eV and T=20~K. The inset shows the region around the $\mathrm{MoS_2}$
$K$ point in high-resolution. An
energy distribution curve (EDC) cut of that data at the $K$-point of $\mathrm{MoS_2}$ (labelled by $K_{MoS_2}$)
is depicted in (b). The
extracted spin-orbit splitting is 144~meV. (c) Fermi-surface map of graphene. The blue dot in the Fermi surface map
denotes the position of the $K$ point of graphene and is labelled by $K_{Gr}$ in (a). We extract a hole density of $1.484 \times10^{13}~\mathrm{cm}^{-2}$. (d) ARPES scan of Li-doped $\mathrm{MoS_2/gr/Ir(111)}$. (e) High
resolution scan close to the Fermi level around the $K$ point of doped
$\mathrm{MoS_2}$ shows the conduction band shifting below the Fermi level upon Li
evaporation. As a guide to the eye we have inserted a parabola shown in red. EDC
cuts through the data are shown to the right of the ARPES
scans in blue. A fit to the data is shown in black.}\label{fig:overview}
\end{figure}

\section{Experimental Results}
\subsection{Structure and electronic properties}
Prior to the analysis of the electronic and optical properties of the \mo~layer, we present in Figure 1 its microscopic, structural and chemical characterization. In
(a), a large scale STM topograph of the \mo~island
layer is shown. The islands rest on the Gr/Ir(111) substrate, which has two
monatomic step edges crossing the topograph horizontally. A large fraction of the substrate is covered by monolayer islands (green arrow in the inset), decorated with small bilayer islands (black arrow). On both, the monolayer and the bilayer, a metallic edge state surrounding the islands can be observed (blue arrow), since the bias voltage lies in the
band gap of the semiconductor \mo. Bright lines, running across the \mo~islands can
be identified as (mirror) twin boundaries (white arrow)~\cite{Jolie18}. The \mo~islands are
extremely clean with a negligible density of defects. Subfigure (b) displays a LEED pattern of the sample, indicating the epitaxial
relation between the substrate and the adlayer. Going from outside to inside, the first order Gr and Ir(111) spots and their associated moir\'e can be seen. Farthest inside,
slightly rotationally broadened first order \mo~diffraction spots indicate a lattice constant of
$(3.13~\pm~0.03)\,\mathrm{\AA}$, in line with the literature~\cite{Young1968}. To probe the chemical properties, XPS was performed. Figure \ref{fig:XPS}(c) compares the Mo~3d core level
of elemental molybdenum in red (produced by evaporating molybdenum onto the
Gr/Ir(111) surface without any source for sulfur) to the grown MoS$_2$/Gr/Ir(111)
structure in green. It can be seen that the Mo~3d core level is shifted to higher
binding energy by 0.95~eV. This shift is in line with earlier observations of
MoS$_2$ grown on a gold substrate~\cite{Bruix2015}. In these previous results, a splitting of the Mo~3d core level was observed into three components (low binding-energy, mid binding-energy and high binding-energy component)~\cite{Bruix2015}. In comparison, our Mo~3d peak lacks the reported low binding-energy and mid binding-energy components which are attributed to metallic Mo and Mo on the edge of a flake. For the low binding-energy peak, we attribute this to the fact that all available Mo was used up in the reaction to form MoS$_2$ and no elemental Mo is left over. The absence of the mid binding-energy component can be explained by the large island size achieved in this work. This increases the ``bulk'' versus the edge contribution to a point where the edge contribution is negligible. The sulfur 2p peak is shown in Figure~\ref{fig:XPS}(d). Our analysis confirms the growth of crystalline MoS$_2$ and the absence of
amorphous MoS$_3$~\cite{Casalongue2014}. Results of MoS$_2$ grown on gold show an
asymmetry in the S~2p peak~\cite{Bana2018} compared to the present work. This can be
explained by the influence of the gold substrate on the lower sulfur layer. This
asymmetry is not visible for MoS$_2$/Gr/Ir(111), suggesting a negligible influence
of the Gr/Ir(111) substrate on the lower sulfide layer and thus a weak interaction
of the substrate with the grown MoS$_2$ islands. As we will discuss later, this weak
interaction is key to observing PL. The C~1s peak of the graphene layer is shown in
Figure \ref{fig:XPS}(e).

Figure \ref{fig:overview} shows angle-resolved photoemission spectroscopy (ARPES)
results of the same system. An overview scan depicting the bands
of graphene, $\mathrm{MoS_2}$ and the Ir substrate is shown in Figure
\ref{fig:overview}(a). The $K$ point of graphene is at
$\sim$1.7~$\rm\AA^{-1}$ and the  $K$ point of $\mathrm{MoS_2}$ at
$\sim$1.3~$\rm\AA^{-1}$, both are indicated at the top x-axis. The valence band (VB) maximum of $\mathrm{MoS_2}$ appears at the $K$-point consistent with monolayer MoS$_2$. For comparison, bilayer MoS$_2$ (shown in the
supporting information) has the VB maximum at the $\Gamma$ point. By taking the
distance between the VB
maximum of $\mathrm{MoS_2}$ at the $K$-point to the Fermi level (approximately
1.5~eV), it suggests that the Fermi level is closer to the conduction band (CB) than to
the valence band of $\mathrm{MoS_2}$
%it suggests electron doping of $\mathrm{MoS_2}$ because for charge neutral
%$\mathrm{MoS_2}$ the chemical potential would lie in between the VB and the CB
as the measured electronic bandgap is typically below 2.6~eV\cite{Wang2017}.
The splitting of the VB at $K$ due to spin-orbit interaction is clearly seen in the
high resolution scan shown in the inset to Figure \ref{fig:overview}(a). The fit to
the energy distribution curve from a cut through the MoS$_2$ $K$-point is shown in Figure \ref{fig:overview}(b) and reveals a
spin-orbit coupling of 144~meV. Interestingly, graphene is more hole doped
than it was before $\mathrm{MoS_2}$ growth, the Dirac-point binding energy is evaluated to be $E_{Dirac}=-0.25~\mathrm{eV}$ compared to $E_{Dirac}=-0.1~\mathrm{eV}$ in the pristine case~\cite{Kralj2011}. The hole doping can be seen from the ARPES scans and the map
shown in Figures~\ref{fig:overview}(a,c). The fact that hole doping increases
after performing the MoS$_2$ growth on Gr/Ir(111) is also evident from a comparison to other works on
Gr/Ir(111)~\cite{pletikosic09-minigaps,Starodub2011}. Analysis of the Fermi surface yields a hole concentration of $1.48\times 10^{13}~\mathrm{cm}^{-2}$. As we
will see later, this hole doping is also responsible for the shift of the Raman
active $G$ band of Gr. Notably, ARPES does not show any hybridization between
MoS$_2$ and graphene bands which supports the idea that MoS$_2$ is weakly
interacting with Gr.

In order to measure the CB edge using ARPES, we have performed Li
doping which induces an electron transfer from Li to the MoS$_2$ layer thereby
populating its CB. Figures \ref{fig:overview}(d,e) show ARPES spectra of Li doped
MoS$_2$/Gr heterostructures. The doping turns MoS$_2$ into a metal which is
corroborated from
the ARPES observation of a CB at the $K$ point of MoS$_2$ (the CB is visible as a
parabola at the Fermi level in Figure \ref{fig:overview}(e)). Assuming a circular
Fermi surface of Li-doped MoS$_2$, we estimate an electron concentration of $3.2\times 10^{13}$
per cm$^2$.

 The VB shifts down in energy and broadens but is otherwise unchanged. A Li induced phase transition in
MoS$_2$ has been predicted theoretically~\cite{NasrEsfahani2015,Zheng2016} and experimentally
reported in Li intercalated quantum dots~\cite{Chen2018}. Interestingly, in the present system we
do not observe a structural phase transition of MoS$_2$ to a 1T (or 1T') phase which would be
visible as a different band structure in the ARPES
measurements~\cite{He2016,Zheng2016}. %**We ascribe the absence of a 1T phase thus to the epitaxial bilayer nature of the MoS$_2$ graphene heterostructure.**
An energy distribution curve (EDC) through the $K$ point yields peaks at 75~meV (CB
minimum) and 2.05~eV (upper VB maximum). Their difference is equal to 1.975~eV. This
value is similar to what has been measured in potassium intercalated monolayer
MoS$_2$ on bulk MoS$_2$ where
1.86~eV was found~\cite{Eknapakul2014}. In the next section we compare the
obtained VB-CB separation to the energy of the PL to estimate a lower bound of the
exciton binding
energy.

\begin{figure}[htb!]
\centering
\includegraphics[width=\textwidth]{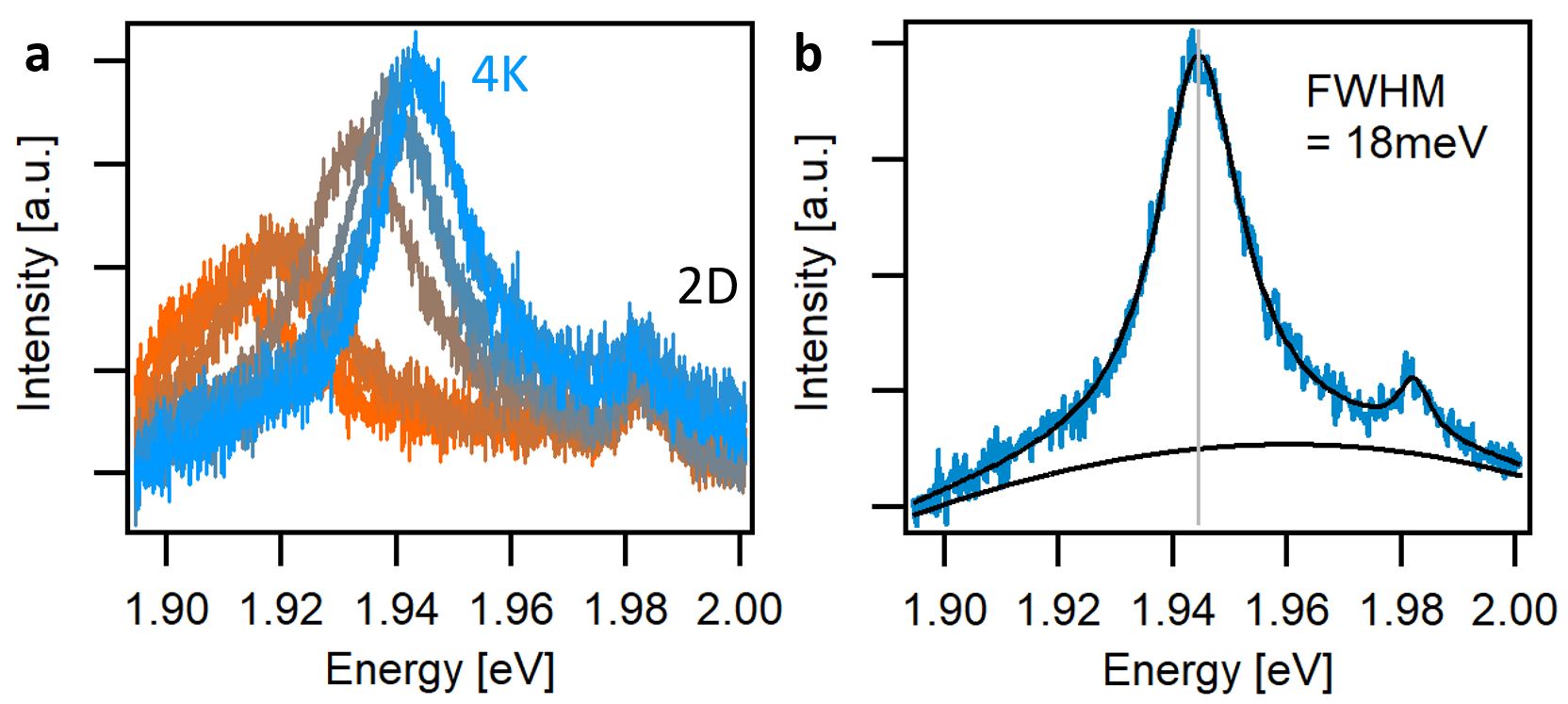}
\caption{a) Ultra-high vacuum photoluminescence (PL)  measurements of MoS$_2$/Gr/Ir(111)
during cooldown to 4~K (orange to blue color) show the expected shift with
temperature. While the PL gets sharper with cooling due to longer lifetimes of the
excitons, the area under the curve stays constant. The data was normalized to the
intensity of the graphene 2D peak at around 1.98~eV. b) At 4~K a Lorenz-fit to the
data using a cubic-spline background determines the peak location at 1.945~eV with a
full width at half maximum (FWHM) of $18~\mathrm{meV}$}. \label{fig:PL}
\end{figure}

\subsection{Luminescent properties}
Samples prepared and characterized in this way have then been transferred without exposure to air to an ultra-high-vacuum (UHV) PL/Raman system~\cite{Grueneis2017}.
Despite the MoS$_2$ islands are grown on a metallic substrate we were able to detect PL at low temperatures. Figure \ref{fig:PL}(a) shows the PL spectra as the sample temperature is lowered. 
Besides the peak that originates from the PL a second order 2D Raman peak from graphene is seen slightly above 1.98~eV. A shift towards higher energy and a narrowing of the linewidth can be observed with decreasing temperature for the PL related peak while the Raman peak of graphene is not shifting. The area under the PL is temperature independent suggesting that the peak becoming more prominent is due to the reduced FWHM at low temperatures.
Figure \ref{fig:PL}(b) shows the PL spectrum recorded at 4~K with a maximum at $E=1.945$~eV together with a lineshape analysis. The narrow width of 18~meV of the PL points towards a long excitonic lifetime at low temperatures. 

Next, we consider the relation between the CB-VB separation from ARPES of doped MoS$_2$ (1.976~eV) and  the PL peak (1.945~eV). Naively (assuming that the Li doping does not affect the band gap value) one might expect that the difference between these two values (30~meV) is equal to the exciton binding energy. However, considering that the bandgap is related to the dielectric function and that doping leads to better screening, we expect a decrease of the bandgap. This has been observed for carbon nanotubes~\cite{Spataru_PRL1010,Hartleb_ACSN15} and graphene nanoribbons~\cite{Senkovskiy_AELM} and theoretically calculated for TMDCs~\cite{Gao2016,Yufeng_Li_PRL2015}. According to quasiparticle calculations, the band gap renormalization due to doping is expected to be the dominant factor that needs to be considered for the determination of exciton binding energies out of such an experiment.
For example, for the present carrier concentration of $3.2\times 10^{13}$ carriers per cm$^{2}$ a band gap reduction by 450~meV is predicted~\cite{Yufeng_Li_PRL2015}. Ignoring this effect would therefore only yield a lower bound of the exciton binding energy. However, if we include the calculated reduction of the ARPES band gap by doping (450~meV), we can estimate an exciton binding energy of about 480~meV. Indeed, this value is very similar to related experiments. Ugeda et al. found an exciton binding energy of 550~meV for MoSe$_2$ on bilayer graphene on 6H-SiC(0001) by comparing PL and STS data~\cite{Ugeda2014}. A recent study combining ARPES and inverse photoemission of the MoS$_2$/Au system by Park et al. found an exciton binding energy of 90~meV~\cite{Park2018a}. This value is considerably lower because of better screening on Au and highlights the important role of the dielectric environment.  Furthermore, a decrease in the band gap upon photodoping has also been observed~\cite{Pogna2016}.

The appearance of PL is surprising because one would expect exciton quenching by the
graphene or the metallic substrate by either F\"orster or Dexter transfer
processes~\cite{Froehlicher2018}. Electroluminescence of monolayer $\mathrm{MoS_2}$
on a gold surface has been observed previously~\cite{Krane2016} by tunneling
electrons directly into the MoS$_2$ via an STM tip. Experimentally it is known that
the interaction between graphene and MoS$_2$ or semiconducting
quantum dots results in luminescence
quenching~\cite{Froehlicher2018,Gaudreau2013a,Federspiel2015}. To the best of our
knowledge there is no theoretical study of the mechanism of exciton quenching in the
present system. However,  a theoretical study of exciton quenching of luminescent
molecules on graphene~\cite{Malic2014} suggests that both F\"orster and Dexter
processes are relevant and graphene is an efficient energy sink. We speculate that
the same is true for the present system.

The efficiency of luminescence quenching in exfoliated
MoS$_2$/Gr heterostructures is reduced by the intercalation of adsorbates into the interface~\cite{Froehlicher2018}. In the present case however, we can rule out such effects because we keep the sample always in either N$_2$ or high vacuum (samples were carried from the growth chamber to the UHV PL/Raman system in a vacuum suitcase or a vacuum tight N$_2$ container). The transferred samples still show a LEED pattern and the apparent height of the MoS$_2$ islands in STM is
unchanged. Therefore, we believe that intercalation of adsorbates into the MoS$_2$/Gr interface is not responsible for the appearance of PL. Instead, we suspect that the key for PL observation is the relatively weak graphene-MoS$_2$ interaction as already discussed in the context of XPS and ARPES data analysis. To learn more about this interaction, we show temperature dependent Raman spectroscopy data taken inside UHV in the next section.

\begin{table}
\centering
\caption[]{\label{tab:raman}The frequency of the $E_{2g}$ and $A_{1g}$ is denoted by
$\omega$, $\chi$ denotes the change of phonon frequency with temperature and
$\gamma$ the Gr\"uneisen parameter.\\
$^a$ This work MoS$_2$/Gr/Ir(111) measured at RT.\\
$^b$ Lee et al. exfoliated MoS$_2$ on SiO$_2$~\cite{Lee2010}.\\
$^c$ Rice et al.~\cite{Rice2013} have determined $\omega$ on a polymer and $\gamma$
from four point bending.\\
$^d$ Sahoo et al.~\cite{Sahoo2013} have determined $\omega$ on SiO$_2$ and $\chi$
between 80~K$-$473~K.\\
$^e$ Najemaei et al.~\cite{Najmaei2013} have determined $\chi$ in the range
300~K$-$500~K.\\
$^f$ Yan et al.~\cite{Yan2014-acs} have performed measurements of $\chi$ for
suspended monolayers and found that $\chi$ for sapphire supported monolayer is
similar.\\
$^g$  Sugai et al.~\cite{Sugai1982}}
\begin{tabular}{cccccc}
\mr
Mode & $\omega$[cm$^{-1}$] & $\chi$[cm$^{-1}$/K] & $\gamma$ (ML) & $\gamma$ (bulk) \\
\mr
A$_{\mathrm{1g}}$ & 405.1$^a$, 403.0$^b$, 402.4$^c$, 408.4$^d$, 405.0$^e$ &
-0.013$^f$,-0.0123$^d$,-0.0143$^e$ & 0.21$^c$ & 0.21$^g$\\
$\mathrm{E_{2g}}$ & 384.1$^a$, 384.5$^b$, 385.3$^c$, 382.6$^d$, 385.0$^e$ &
-0.011$^f$, -0.0132$^d$, -0.0179$^e$ & 0.65$^c$ & 0.42$^g$ \\ 
\br
\end{tabular}\end{table}

\begin{figure}[htb!]
\centering
\includegraphics[width=\textwidth]{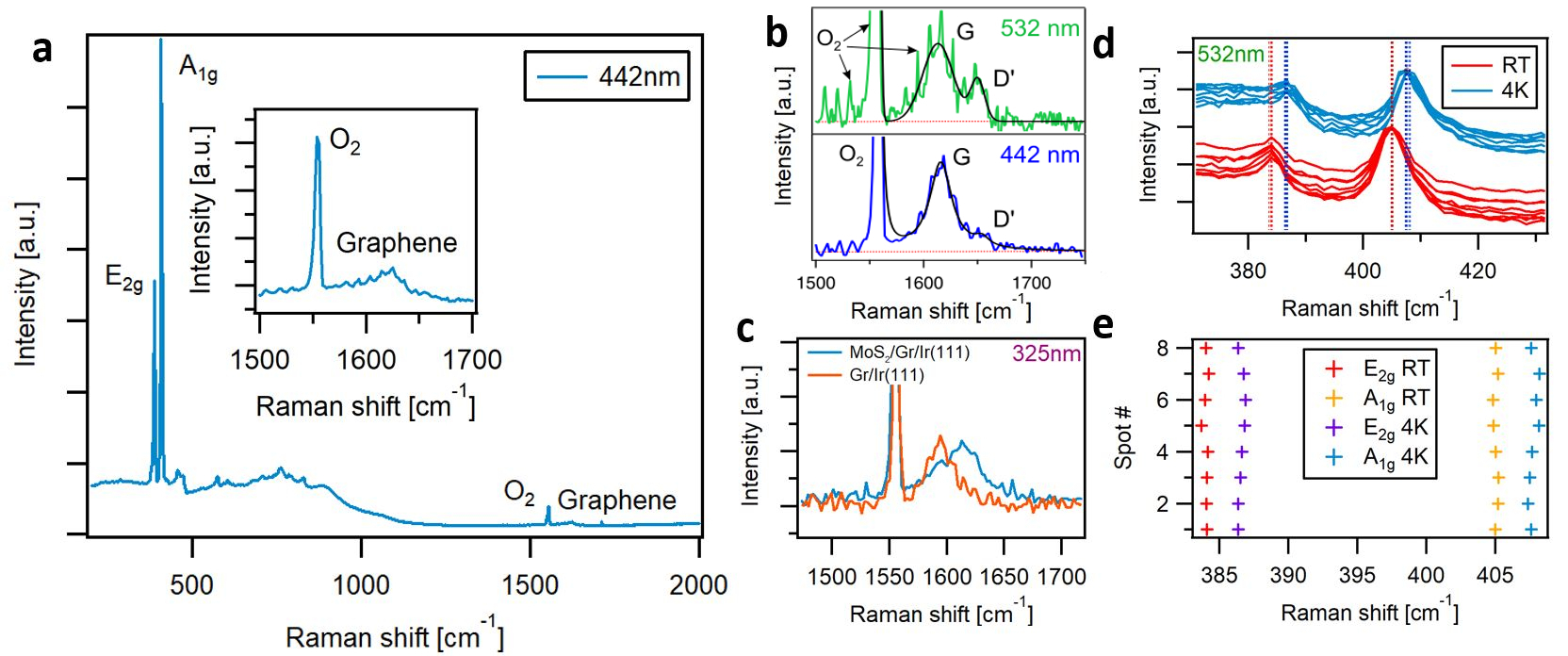}
\centering
\caption{Ultra-high vacuum Raman spectroscopy of $\mathrm{MoS_2/Gr/Ir(111)}$.
(a) Overview scan at 4K using 442~nm  wavelength laser. The A$_{\rm 1g}$ and
E$_{\rm 2g}$ Raman active modes of $\mathrm{MoS_2}$ are very strong compared to the
graphene related modes. The peak at $1555~\mathrm{cm^{-1}}$ is due to the vibrations of molecular $\mathrm{O_2}$ in the beam path outside of the UHV chamber. The inset shows the region around $1600~\mathrm{cm^{-1}}$ with the graphene $G$ peak.
(b) $D'$ peak measurements at room temperature for 532~nm and 442~nm excitation show that the $D'$
intensity is lower for the smaller wavelength. The small peaks around the O$_2$ vibrational peak are the rotational states of O$_2$ which are more pronounced for visible light excitation than UV. 
(c) Comparison of the room temperature $G$-peak position of $\mathrm{MoS_2}$/Gr/Ir(111) and
Gr/Ir(111)  shows a shift from
1593.4~$\mathrm{cm^{-1}}$ to 1613.1~$\mathrm{cm^{-1}}$. (d) Temperature dependence
of the $\mathrm{MoS_2}$ Raman modes using a 532~nm laser at eight
different spots on the sample. (e) Peak positions from Lorentzian fits to the data
shown in (d). Their avarage values at 4~K are  $\omega_{\mathrm{E_{2g}}}=386.6~\mathrm{cm^{-1}}$ and $\omega_{\mathrm{A_{1g}}}=407.8~\mathrm{cm^{-1}}$ and at RT
the average values  $\omega_{\mathrm{E_{2g}}}=384.1~\mathrm{cm^{-1}}$ and
$\omega_{\mathrm{A_{1g}}}=405.1~\mathrm{cm^{-1}}$.  } \label{fig:raman}
\end{figure}

\subsection{Vibrational properties and strain}
Raman spectra have been taken in the same experimental setup as the PL measurements inside a vacuum better than $2\times 10^{-10}$~mbar (see Methods). Figure \ref{fig:raman}(a) shows
an overview Raman spectrum taken at 4~K using a 442~nm excitation. The $\mathrm{MoS_2}$ related phonons with A$_{\rm 1g}$ and E$_{\rm 2g}$ symmetry are
strong in intensity compared to graphene (see inset of Figure \ref{fig:raman}(a))
and have a
splitting of 21~$\mathrm{cm^{-1}}$. A comparison of $G$ band Raman spectra for
442~nm and 532~nm excitation is shown in Figure \ref{fig:raman}(b). A shoulder at
$\sim$1650~$\mathrm{cm^{-1}}$ can be identified which is attributed to the $D'$ band 
because its intensity is changing with laser energy~\cite{cancado06-dprime,pimenta07-review}, as one can see from Figure 4(b).
The appearance of the $D'$ Raman band is ascribed to translational symmetry breaking by the
$\mathrm{MoS_2}$ islands (see the STM image in Figure \ref{fig:XPS}(a)) which act as
scattering centers for graphene electrons. Importantly, the graphene $G$ band prior to $\mathrm{MoS_2}$  synthesis is
not visible by 442~nm and 532~nm excitation.~\cite{martin18} However, it can be
detected using UV excitation (325nm). The comparison of UV Raman spectra for
Gr/Ir(111) and the $\mathrm{MoS_2}$/Gr/Ir(111) heterostructure are shown
in Figure \ref{fig:raman}(c). A shift of $G$ band position from 1593~$\rm cm^{-1}$
(Gr/Ir) to 1613~$\rm cm^{-1}$ ($\mathrm{MoS_2}$/Gr/Ir) can be seen. The frequency upshift by 20~$\rm cm^{-1}$ can be explained by two effects that take place after growth of MoS$_2$. First, graphene becomes p-doped (that we have analyzed by ARPES). The $G$ band frequency upshift upon p-doping has been reported in the literature.~\cite{lazerri06-kohn,Das2008,fengwang11-quantum,martin18} For the observed hole concentration of $1.48\times 10^{13}~\mathrm{cm}^{-2}$ an upshift from the position of charge neutral graphene by $\sim$18~$\mathrm{cm}^{-1}$ is predicted.~\cite{Das2008} Neglecting the small initial p-doping of Gr/Ir(111), this is in very good agreement to the observed 20~$\rm cm^{-1}$ upshift. Notably, the precise value of the upshift depends also on the substrate and other works report values in the range of $\sim$5-10~$\rm cm^{-1}$~(Refs.~\cite{lazerri06-kohn,fengwang11-quantum,martin18}). Second, we believe that, after MoS$_2$ growth, Gr on Ir(111) becomes flatter which leads to compressive strain in Gr. This is corroborated by the fact that the Gr/Ir(111) moir\'e spots in the LEED pattern become weaker after MoS$_2$ growth. The wavyness of the moir\'e can help to relax some of the strain in the Gr/Ir(111) system. However, after MoS$_2$ growth, as Gr becomes flatter, it also acquires compressive strain which is known to cause an upshift in the $G$ band frequency~\cite{lazerri06-kohn,ferrari09-strain}.\newline

Let us now move to the investigation of temperature induced strain in the
heterostructure.  Raman spectroscopy is a well suited tool to investigate the change
of bond length due to strain via the frequency change of Raman active vibrations.
The information of that frequency change versus temperature yields information on
how strongly bonded graphene and $\mathrm{MoS_2}$ are to each other and to the
substrate. For example, if both layers would follow the thermal expansion of the Ir
substrate, we can assume that they are strongly bonded to each other. For graphene
which is in direct contact to the Ir surface and fully covering it, one might expect
that the C-C bond length follows the thermal expansion of the bulk Ir.
However, the situation of MoS$_2$ is less obvious because it is not in direct
contact to the Ir and  not a complete monolayer which can make it easier to maintain
a thermal expansion coefficient of its own.
Figure \ref{fig:raman}(d) depicts scans at several spots on the sample performed at
RT and at 4~K. It can be seen that, upon cooling the phonons harden by 2.7~$\rm
cm^{-1}$ (the A$_{\rm 1g}$ mode) and by 2.6~$\rm cm^{-1}$ (the E$_{\rm 2g}$ mode).
The temperature dependent phonon frequency is phenomenologically described as
$\Delta \omega=\chi\Delta T$ where $\Delta\omega$ is the frequency shift and $\chi$ is a
phonon shift per Kelvin. For $\mathrm{MoS_2}$ there is a consensus in the literature that $\chi\sim -0.01$~$\rm cm^{-1}/K$~\cite{Sahoo2013,Su2014,Yang2016} (see Table 1).
Interestingly, this number is largely independent of the substrate and holds also for freestanding layers~\cite{Yan2014-acs}. It is approximately the same for both
A$_{\rm 1g}$ and E$_{\rm 2g}$ phonon modes. Plugging in $\Delta T\sim$290~K, we
would expect a shift by $\Delta\omega=$3.5~$\rm cm^{-1}$ for freestanding MoS$_2$ which is close to
explaining the experimental value, but not in perfect agreement with the measured
2.6~cm$^{-1}$.
Considering alternative scenarios, the other extremum is strongly substrate bound
$\mathrm{MoS_2}$. In this case, the phonon shift is dictated by the temperature
induced substrate strain $\epsilon$ caused by the change of the substrate lattice
parameter, to which MoS$_2$ would be pinned. This strain can be derived from the
linear thermal expansion coefficient of
iridium~\cite{White1970-iridium_low,Halvorson1972-iridium}.
We proceed by first applying this analysis to the graphene $G$ band and then to
MoS$_2$. The strain $\epsilon$ and the phonon shift $\Delta\omega$ are linked to
each other via the Gr\"uneisen parameter $\gamma$ and the phonon mode degeneracy $n$
as $\Delta\omega=\epsilon n\gamma\omega_0$. Here $\omega_0$ is the phonon frequency
of the unstrained system. The temperature induced strain $\epsilon$ for the Ir
substrate yields $\epsilon=0.134\%$. Plugging this into the above equation for the
graphene $G$ mode with E$_{\rm 2g}$ symmetry and using $\omega_0=1593.2$~$\rm
cm^{-1}$, $n=2$ $\gamma=2$,  we obtain $\omega=1604.1$~$\rm cm^{-1}$
(in Ref.~\cite{martin18} this analysis has been performed for the first time for Gr/Ir(111) and more details can be found there).
Importantly, the temperature dependent upshift in $\omega$ upon cooling for the $G$ band is in good agreement to the experiment. This implies that
graphene is pinned to the Ir substrate~\cite{martin18}.

Performing the same estimation for $\mathrm{MoS_2}$ we try to obtain a value for the temperature dependent phonon energy shift. The
Gr\"uneisen parameters of monolayer $\mathrm{MoS_2}$ are reported in the literature
as $\gamma_{\mathrm{A_{1g}}}=0.21$ and $\gamma_{\mathrm{E_{2g}}}=0.65$ (Ref.~\cite{Rice2013}). If we now
apply the above formula, assuming that $\epsilon$ is that of the strained Ir substrate, we find $\Delta\omega_{\mathrm{A_{1g}}}=0.11$~$\mathrm{cm}^{-1}$ and $\Delta\omega_{\mathrm{E_{2g}}}=0.67$~$\mathrm{cm}^{-1}$. This does not agree with experiment at all.
Notably, also using the Gr\"uneisen parameter of bulk MoS$_2$ ($\gamma_{\mathrm{A_{1g}}}=0.21$ 
and  $\gamma_{\mathrm{E_{2g}}}=0.42$ from~Ref.~\cite{Sugai1982}) would not improve agreement. We thus conclude that $\mathrm{MoS_2}$ does not follow the thermal expansion of Ir and its behaviour is better described by the expansion expected for a freestanding monolayer. Graphene, however, is stronger interacting with the Ir substrate and its Raman shift as a function of temperature can be fully understood by the thermal expansion of the substrate.

\section{Conclusion and Outlook}
We have characterized the epitaxially grown MoS$_2$/Gr/Ir(111) system combining XPS,
ARPES, Raman and PL measurements. STM, LEED and XPS confirm the good quality of our
grown samples. We have observed a PL-signal with small FWHM suggesting a long excitonic
lifetime. This surprising result is the first clear observation of
photoluminescence of epitaxially grown MoS$_2$ on a metallic substrate. The absence
of the expected quenching of the PL intensity on a metallic surface can potentially
be explained by a weak interaction between the epitaxial MoS$_2$ and the substrate
as is suggested by our XPS, ARPES and temperature dependent Raman measurements.
Using Li deposition, we induced doping of MoS$_2$ into a degenerate semiconductor
and obtained from the analysis of ARPES data the band gap of Li-doped MoS$_2$. Using theoretical calculations on the band gap renormalization due to doping, we estimate an exciton binding energy of 480~meV. Our
results suggest that the MoS$_2$-islands are only weakly interacting with the Gr/Ir
surface which could explain the absence of quenching, but the microscopic mechanisms
are still unclear. Theoretical calculations for the Dexter- and F\"orster-type
energy transfer from the islands into the graphene substrate are thus needed to
quantitatively explain the observed PL. With this background it would be interesting
to grow MoS$_2$ on hexagonal boron nitride (h-BN) using the same method as used for this work and compare FWHM
and intensity of the PL. Indeed, previous experiments on h-BN capped
MoS$_2$~\cite{Cadiz2017} have shown an increase in the PL intensity upon h-BN
encapsulation. Similarly, it was shown that chemical treatment of $\mathrm{MoS_2}$ flakes via an organic superacid increased PL quantum yield to near unity~\cite{Amani2015}, similar treatment of epitaxially grown MoS$_2$ monolayers might increase PL intensity even more.
Additionally a transfer of MoS$_2$ islands grown on Gr/Ir(111) onto different substrates could help to understand the effects of the substrate on the luminescent
properties.

\section{Appendix / supporting information}
\begin{figure}[htb!]
\centering
\includegraphics[width=8cm]{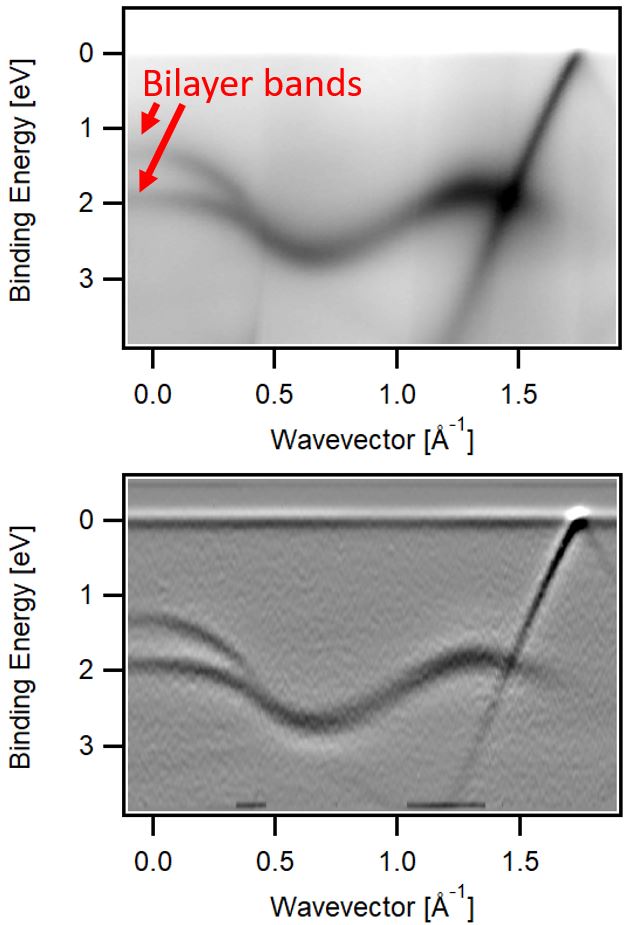}
\caption{ARPES of bilayer MoS$_2$/Gr. Upper panel: raw data, lower panel: second
derivative for enhancing spectral features.}\label{fig:bilayer}
\end{figure}

%\begin{figure}[htb!]
%\includegraphics[width=7cm]{./figures/bilayerRaman.jpg}
%\includegraphics[width=7cm]{./figures/bilayerPL.jpg}
%\caption{Raman and PL of bilayer MoS2/graphene. ** Temperature  ** Peak assignement
%of Raman peaks is wrong , 2D should be 2G and 2G is an unknown phonon **}
%\end{figure}

\subsection{ARPES of bilayer MoS$_2$}
Figure \ref{fig:bilayer} depicts ARPES spectra of bilayer  MoS$_2$ that has been grown by doubling the deposited amount of Mo. This resulted in 1.4 monolayers (ML) of MoS$_2$ but growth conditions were specifically tuned to induce bilayer growth via
sulfur pressure in the chamber and cycled growth (see Methods section). It is clear from ARPES that the VB maximum is not at the $K$ point but at the $\Gamma$ point (note the splitting of the band at $\Gamma$ into two subbands, one with a higher binding energy of approximately 1.9~eV and one with a lower binding energy of
approximately 1.3~eV).

%\color{red}
%\subsection{Open questions}
%\begin{itemize}
%\item How much Dexter/F\"orster transfer from MoS2 to graphene do we have?
%\item What would be the rater of F\"orster transfer if MoS2 was on hBN (ongoing
%expt) ? 
%\item What is the calculated Exciton binding energy of MoS2 / gr / Ir ? Does it fit
%to expt?
%\item Inverse photoemission + PL suggest an excitonic BE similar to
%ours~\cite{Park2018a}. Does it mean that Li is not modifying CB-VB separation?
%\item Does theory support the idea of ``freestanding'' MoS2 that is indicated by
%the temperature dependent phonon shift (Raman section) ?
%\item What is the peak in the bilayer PL at 1.9~eV ?
%\item Why is MoS2 negatively doped ? 
%\end{itemize}
%\color{black}

\section{Methods}
\subsection{X-ray photoemission spectroscopy}
XPS was performed at the German-Russian beamline (RGBL) of the HZB BESSY II
synchrotron in Berlin (Germany) with a beam energy of 650~eV and pass energy of 20~eV
in a normal emission geometry. The MoS$_2$/Gr/Ir(111) samples were prepared in-situ
and
measured in a vacuum better than $5\times 10^{-10}$~mbar.

\subsection{Angle-resolved photoemission spectroscopy}
ARPES was performed at the BaDElPh beamline~\cite{Petaccia2009} of the Elettra
synchrotron in Trieste (Italy) with linear s- and p- polarisation at $h\nu=31$~eV at temperatures of 20~K. The MoS$_2$/Gr/Ir(111) samples were prepared in-situ and measured in a vacuum better than $5\times 10^{-11}$~mbar. Li deposition was carried out in an ultra-high vacuum (UHV) chamber from SAES getters with the sample at 20~K.
We performed stepwise evaporation of Li which we monitored by ARPES measurements of the band structure. Li evaporation was stopped after the desired doping level was reached.

\subsection{Scanning Tunneling Spectroscopy and Microscopy}
Scanning tunneling microscopy was conducted in a home built variable temperature
STM apparatus in Cologne at a base pressure below $8 \times 10^{-11}$~mbar. For image
processing the software WSxM was used~\cite{Horcas2007}.

\subsection{Growth}
We employ molecular beam epitaxy via a two-step process~\cite{Hall2018}: In the
first step, with the sample held at room temperature, Mo is evaporated at a rate of
$\approx1.4\times10^{16}$~atoms m$^{-2}$ s$^{-1}$ into a S~background pressure of
$p\approx5\times10^{-9}$~mbar onto Gr/Ir(111). The elemental S~background atmosphere
is achieved by heating a pyrite (FeS$_2$) filled crucible to $\approx$~500~K. During
the second step, the sample is annealed for 300\,s at $T=1050$~K in a S pressure of
$p\approx2\times10^{-9}$~mbar. These two steps constitute one growth cycle.\\

To obtain a \mo~layer with orientation epitaxy even for coverages beyond 0.4\,ML the
total coverage was deposited in subsequent growth cycles each yielding a coverage of
$\approx~0.35$~ML~\mo. Using this technique we realized two cycle \mo~samples
(nominal coverage 0.7~ML) and four cycle \mo~samples (nominal coverage 1.4~ML).\\

\subsection{Ultra-high Vacuum Raman and photoluminescence spectroscopy}
UHV Raman measurements were performed in the back-scattering geometry using
commercial Raman systems (Renishaw) integrated in a homebuilt optical
chamber~\cite{Grueneis2017}, where the exciting and Raman scattered light were
coupled into the vacuum using a 50x long-working distance microscope objective with
an NA of $\sim$0.4 and a focal distance of 20.5~mm for lasers with wavelength
442~nm and 532~nm. For the UV laser, a UV compatible microscope objective has
been used. The $20$x UV objective has a focal distance equal to 13~mm and an
NA=0.32. A sketch of our experimental setup is shown in a previous
work~\cite{Grueneis2017}. Power densities in the range of of 100~kW/cm$^2$ have been employed for all laser energies. The
position of the laser on the sample could be checked by a camera in the laser path.
All spectra have been calibrated in position and intensity to the O$_2$ vibration at
1555~cm$^{-1}$~\cite{Faris1997-oxygen}. 

\section{Acknowledgements}
N.E., J.H., A.H., T.M. and A.G. acknowledge support through the CRC 1238 within
project A01. N.E., B.V.S., M.H. and A.G. acknowledge the ERC grant no. 648589
'SUPER-2D' and funding from DFG projects GR 3708/2-1. A.G. acknowledges INST
216/808-1 FUGG and support from the ``Quantum Matter and Materials'' (QM2)
initiative. N.E., M.H, J.L and A.G. acknowledge support from
CALIPSOplus and CERIC-ERIC for their stay at ELETTRA sincrotrone.

\section*{References}
\bibliographystyle{unsrt}

\bibliography{MoS2_Paper}

\end{document}